\documentclass[12pt]{article}
\begin{document}

\title{\bf Perturbed Self-Similar Massless Scalar Field in Spherically Symmetric
Spaceimes}

\author{M. Sharif \thanks{e-mail: msharif@math.pu.edu.pk}
\\ Department of Mathematics, University of the Punjab,\\
Quaid-e-Azam Campus Lahore-54590, PAKISTAN.}

\date{}

\maketitle

\begin{abstract}
In this paper, we investigate the linear perturbations of the
spherically symmetric spacetimes with kinematic self-similarity of
the second kind. The massless scalar field equations are solved
which yield the background and an exact solutions for the perturbed
equations. We discuss the boundary conditions of the resulting
perturbed solutions. The possible perturbation modes turn out to be
stable as well as unstable. The analysis leads to the conclusion
that there does not exist any critical solution.
\end{abstract}


\date{}

\section{Introduction}

In General Relativity, gravitational collapse of a realistic body is
one of the important problems. At the threshold of black hole
formation, the matter is just about to form a black hole and an
infinitesimally small perturbation can either cause the matter to
disperse to infinity or to form a black hole. The dynamics close to
the threshold exhibits interesting behavior such as power law
scaling of length scales, self-similarity of the solutions and
universality. A typical setup is to take an initial matter
distribution parameterized by a single parameter. The critical
solution is self-similar, i.e., it repeats itself on ever decreasing
length scales. The fluid density and pressure increase by many order
of magnitudes and in some cases, the fluid velocity becomes
extremely relativistic. Another way of investigating the critical
collapse is to use the fact that the critical solution is
continuously self-similar. Self-similarity leads to the set of
partial differential equations into a set of ordinary differential
equations that can then be solved quite easily numerically to a very
high precision. One then performs a perturbation analysis around the
critical solution to determine the scaling exponent.

Critical phenomena in gravitational collapse were discovered by
Choptuik [1,2] in the spherically symmetric collapse of massless
scalar field. Since then the phenomenon has been observed in a
variety of matter sources. There are two steps to find critical
solutions: Firstly, find a generic family (or families) of
solutions, defined by a parameter, say $p$ such that when $p>p^*$
the collapse forms black holes and when $p<p^*$, it does not.
Secondly, the perturbations of the solution $p=p^*$ are performed to
investigate the spectrum of their modes. If the solution has only
one unstable mode, then this solution is a critical solution and the
exponent $\gamma$ is given by
\begin{equation}
\gamma=\frac{1}{|\sigma|},
\end{equation}
where $\sigma$ is the unstable mode [3].

Garfinkle [4] found a class, say, $S[n]$, of exact solutions to the
Einstein massless scalar field equations in $(2+1)$-dimensions. He
showed that in the strong field regime the $n=4$ solution fits very
well with the numerical critical solution found by Pretorious and
Choptuik [5]. Later, Garfinkle and Gundlach [6] studied their linear
perturbations and found that only the solution with $n=2$ has one
unstable mode, while the one with $n=4$ has three. They further
required that no matter field should come out of the already formed
black holes. This additional condition seems physically quite
reasonable and has been widely used in the investigation of black
hole perturbations [7]. Hirschmann et al. [8] systematically studied
the critical gravitational collapse of a scalar field. They surveyed
all the analytic, continuously self-similar solutions and also
examined their perturbations considering their global structure.
Clement and Fabbri [9] investigated analytical treatment of critical
collapse in $2+1$-dimensional AdS spacetime. Cavaglia et.al. [10]
analysed approximately self-similar critical collapse in
$2+1$-dimensions.

Miguelote et al. [11] studied the gravitational collapse of
self-similar perfect fluid in $2+1$-dimensional spacetimes with
circular symmetry. They also studied the linear perturbations of
homothetic self-similar stiff fluid solutions. Frolov [12] studied
the perturbations of the continuously self-similar critical solution
of the gravitational collapse of a massless scalar field. Brandt et
al. [13] studied the gravitational collapse of spherically symmetric
perfect fluid with kinematic self-similarity. They also studied the
solutions of the Einstein field equations found by Benoit and Coley
[14] and concluded that some of the solutions represent
gravitational collapse. Wang [15] studied the critical collapse of a
cylindrical symmetric scalar field. He introduced the notion of
homothetic self-similarity to four-dimensional spacetimes and then
presented a class of exact solutions to the Einstein massless scalar
field equations. Wang et al. [16] studied plane symmetric
self-similar solutions to Einstein's four-dimensional theory of
gravity and explored the local and gravitational conditions.

Recently, Chan et al. [17] investigated the solution of the Einstein
massless scalar field equations with kinematic self-similarity of
the second kind in the (2+1)-dimensional spacetimes with circular
symmetry. They discussed their local and global properties and also
found that some of these solutions represent gravitational collapse
of the scalar field. The same authors [18] studied the linear
perturbations of the (2+1)-dimensional circularly symmetric solution
with kinematic self-similarity of the second kind. They obtained an
exact solution for the perturbation equations and the possible
perturbation modes and showed that the background solution is a
stable.

In this paper, we investigate the linear perturbations of the
spherically symmetric spacetimes with kinematic self-similarity of
the second kind. We analyze these solutions to see whether the
solution is stable or not. The possible perturbation modes are also
discussed. The paper has been organized as follows. In the next
section, we shall consider the general spherically symmetric
spacetimes and calculate self-similar variable of the second kind.
Section 3 is devoted to the linear perturbation of the field
equations. In sections 4 and 5, we find the possible solutions of
the linear perturbation equations and check the boundary conditions
for the perturbed solutions respectively. Section 6 is focussed for
discussion of the results.

\section{Self-Similarity of the Second Kind}

In this section, we calculate the self-similar variable of the
second kind for spherically symmetric spacetimes. Further, we write
down the non-zero components of the Ricci tensor in terms of the
self-similar variable. The most general form of the spherically
symmetric metric is given by [19]
\begin{equation}
ds^2=e^{2\Phi(t,r)}dt^2-e^{2\Psi(t,r)}dr^2-r^2S^2(t,r)(d\theta^2+
\sin^2\theta d\phi^2),
\end{equation}
where $\Phi(t,r),~\Psi(t,r),~S(t,r)$ are arbitrary functions of
$t,r$. The non-zero components of the Ricci tensor are
\begin{eqnarray}
R_{00}&=&e^{2(\Phi-\Psi)}[\Phi_r(\Phi_r-\Psi_r+\frac{2S_r}{S}+
\frac{2}{r})+\Phi_{rr}]\nonumber\\&-&\frac{2S_{tt}}{S}
+\frac{2\Phi_t S_t}{S}+\Phi_t\Psi_t-\Psi^2_t-\Psi_{tt}, \\
R_{01}&=&\frac{2\Psi_t}{r}+\frac{2\Psi_t S_r}{S}+
\frac{2\Phi_r S_t}{S} -\frac{2S_t}{r S}-\frac{2S_{rt}}{S}, \\
R_{11}&=& e^{2(\Psi-\Phi)}[\Psi_{tt}+
\Psi_t(\Psi_t+\frac{2S_t}{S}-\Phi_t)]-\Phi_{rr}\nonumber\\
&+&\Phi_r\Psi_r-\Phi^2_r-\frac{2S_{rr}}{S}-\frac{4S_r}{r S}
+\frac{2\Psi_r S_r}{S}+\frac{2\Psi_r}{r},\\
R_{22}&=&e^{-2\Phi}(r^2S_t^2+r^2S S_{tt}-r^2S S_t\Phi_t+r^2S
S_t\Psi_t)\nonumber\\&+&e^{-2\Psi} (-S^2-r^2S_r^2-r^2S S_{rr}+r
S^2\Psi_r+r^2S S_r\Psi_r
\nonumber\\&-&4r S S_r-r S^2\Phi_r-r^2S S_r\Phi_r)+1,\\
R_{33}&=&R_{22}\sin^2\theta,
\end{eqnarray}
where the subscripts $t$ and $r$ mean differentiation w.r.t $t$ and
$r$ respectively. We consider here solutions with kinematic
self-similarity of the second kind for which the self-similar
variable $x$ turns out to be
\begin{equation}
x=\ln(\frac{r}{(-t)^{\frac{1}{\alpha}}}),\quad \tau=-\ln(-t),
\end{equation}
or inversely
\begin{equation}
r=e^\frac{\alpha x-\tau}{\alpha}, \quad t=-e^{-\tau},
\end{equation}
where $\alpha$ is a dimensionless constant. The components of the
Ricci tensor in terms of the self-similar variable can be written as
follows
\begin{eqnarray}
R_{00}&=&\frac{e^{2(\Phi-\Psi)}}{r^2}\{\Phi_x(\Phi_x-\Psi_x
+2\frac{S_x}{S}+1)+\Phi_{xx}\}-\frac{1}{\alpha^2
t^2}\{\alpha^2[\Psi_{\tau\tau}\nonumber\\&+&\Psi_\tau
(1+\Psi_\tau-\Phi_\tau)+2\frac{S_{\tau\tau}}{S}+
2\frac{S_\tau}{S}(1-\Phi_\tau)]+\alpha[2\Psi_{\tau x}
+\Psi_\tau(\Psi_x\nonumber\\&-&\Phi_x)+\Psi_x
(\Psi_\tau-\Phi_\tau+1)+ 4\frac{S_{\tau
x}}{S}+2\frac{S_x}{S}(1-\Phi_\tau)-2\frac{S_\tau
\Phi_x}{S}]\nonumber\\&+&[\Psi_{xx}+\Psi_x(\Psi_x-\Phi_x)
+2\frac{S_{xx}}{S}-2\frac{S_x \Phi_x}{S}]\},\\
R_{01}&=&-\frac{2}{\alpha t
r}\{\alpha[\Psi_\tau(1+\frac{S_x}{S})+\frac{S_\tau}{S}
(\Phi_x-1)-\frac{S_{\tau x}}{S}]\nonumber\\&+&\Psi_x
(1+\frac{S_x}{S})+\frac{S_x}{S}(\Phi_x-1)-\frac{S_{xx}}{S}\},\\
R_{11}&=&\frac{e^{2(\Psi-\Phi)}}{\alpha^2
t^2}\{\alpha^2[\Psi_{\tau\tau}+\Psi_\tau(1+\Psi_\tau
-\Phi_\tau+2\frac{S_\tau}{S})]\nonumber\\&+&\alpha[2 \Psi_{\tau
x}+\Psi_x(1+2\Psi_\tau-\Phi_\tau+2\frac{S_\tau}{S})+\Psi_\tau(
-\Phi_x+2\frac{S_x}{S})]\nonumber\\&+&\Psi_{xx}+\Psi_x(\Psi_x-
\Phi_x+2\frac{S_x}{S})\}+\frac{1}{r^2}[\Phi_x(\Psi_x-\Phi_x+1)
\nonumber\\&-&\Phi_{xx}+2\Psi_x(1+\frac{S_x}{S})-
\frac{2}{S}(S_{xx}+S_x)], \\
R_{22}&=&r^2 S^2\{\frac{e^{-2\Phi}}{\alpha^2 t^2
S}[\alpha^2(S_\tau(1+\Psi_\tau-\Phi_\tau+\frac{S_\tau}{S})
+S_{\tau\tau})\nonumber\\&+&\alpha(S_\tau(\Psi_x-\Phi_x
+2\frac{S_\tau}{S})+S_x(1+\Psi_\tau-\Phi_\tau)+2S_{\tau
x})\nonumber\\&+&S_x(\Psi_x-\Phi_x+\frac{S_x}{S})+S_{xx}]
-\frac{e^{-2\Psi}}{r^2}[\frac{1}{S}(S_x(3+\Phi_x\nonumber\\
&-&\Psi_x+\frac{S_x}{S})+S_{xx})+\Phi_x-\Psi_x+1]\}+1, \\
R _{33}&=&R_{22}\sin^2\theta.
\end{eqnarray}

\section{Linear Perturbation of the Field Equations}

This section is devoted to set up the Einstein field equations by
using linear perturbation. For this purpose, we take
\begin{eqnarray}
\Phi(\tau ,x)&=&\Phi_0(x)+\epsilon \Phi_1(x) e^{k\tau},\nonumber\\
\Psi(\tau ,x)&=&\Psi_0(x)+\epsilon \Psi_1(x) e^{k\tau},\nonumber\\
S(\tau ,x)&=&S_0(x)+\epsilon S_1(x) e^{k\tau},\nonumber\\
\phi(\tau ,x)&=&\phi_0(x)+\epsilon \phi_1(x) e^{k\tau},
\end{eqnarray}
where $\epsilon$ is a very small real constant and $k$ is an
arbitrary constant. The quantities with subscripts $0$ and $1$
denote background self-similar solutions and perturbations
respectively. It is understood that there may be many perturbation
modes for different values (possibly complex) of the constant $k$.
The general perturbation will be the sum of these individual modes.
The modes with $Re(k)>0$ grow as $\tau\rightarrow \infty$ and are
referred to as unstable modes while the ones with $Re(k)<0$ decay
and are referred to as stable modes. By definition, critical
solutions will have one and only one unstable mode.

We take the following background solution [18]
\begin{eqnarray}
\Phi_0(x)&=&0, \quad
\Psi_0(x)=-\frac{1}{2}\alpha x,\nonumber\\
S_0(x)&=&\frac{2}{2-\alpha}e^{-\frac{1}{2}\alpha x}, \quad
\phi_0(x)=2q\ln(-t)
\end{eqnarray}
and the apparent horizon is given by
\begin{equation}
r_{AH}(t)=[(2-\alpha)\sqrt{-t}]^{\frac{2}{2-\alpha}}, \quad
\alpha<2,
\end{equation}
where $q=\pm\frac{1}{\sqrt{8}}$. For a massless scalar field $\phi$,
the Einstein field equations are
\begin{equation}
R_{ab}=\kappa\phi_a \phi_b ,\quad a,b=0,1,2.
\end{equation}
Here we choose units such that $\kappa=1$ for the sake of
simplicity. Using Eq.(15) in Eqs.(10)-(14), it follows that
\begin{equation}
R_{ab}=R_{ab}(\tau,x,\epsilon)
\end{equation}
If we take $R_{ab}$ as a function of $\epsilon$ only and expand it
in terms of $\epsilon$, it follows that
\begin{equation}
R_{ab}(\tau,x,\epsilon)=\frac{1}{(-t)^2}\{R_{ab}^{(0)}(x)+ \epsilon
R_{ab}^{(1)}(x)e^{k\tau}+O(\epsilon^2)\}.
\end{equation}
The non-vanishing components of the Ricci tensor upto first order in
$\epsilon$ are
\begin{eqnarray}
R_{00}^{(1)}(x)&=&e^{2(\Phi_0-\Psi_0-x+\frac{\tau}{\alpha})}
\{\Phi_0'(2\Phi_1'-\Psi_1')-\Phi_1'\Psi_0'+\Phi_1''\nonumber\\
&+&\Phi_1'+2(\Phi_1-\Psi_1)[\Phi_0'(\Phi_0'-\Psi_0')+\Phi_0''
+\Phi_0']+\frac{1}{S_0}[-2\frac{S_0'S_1}{S_0}\Phi_0'\nonumber\\
&+&4(\Phi_1-\Psi_1)\Phi_0'S_0'+2\Phi_0'S_1'+2\Phi_1'S_0']\}
+\frac{e^{2\tau}}{\alpha^2}\{\Phi_0'(\alpha k
\Psi_1\nonumber\\&+&\Psi_1')+\Psi_0'(\alpha k
\Phi_1+\Phi_1')-2\Psi_0'(\alpha k \Psi_1+\Psi_1')-\alpha^2
k^2\Psi_1\nonumber\\&-&2\alpha k \Psi_1'-\Psi_1''-\alpha^2
k^2\Psi_1-\alpha \Psi_1'+\frac{1}{S_0}[2\Phi_0'(\alpha k
S_1+S_1')\nonumber\\&+& 2 S_0'(\alpha k \Phi_1+\Phi_1')- 2
\alpha^2k^2 S_1-2\alpha^2 k S_1-4\alpha k S_1'-2
S_1''\nonumber\\&-&2\alpha^2 k S_1-2\alpha S_1'
-\frac{S_1}{S_0}(2S_0'\Phi_0'-2S_0''- 2\alpha S_0')]\},\\
R_{01}^{(1)}(x)&=&-2\frac{e^{\frac{\alpha+1}{\alpha}\tau-x}} {\alpha
S_0}[-\frac{S_1}{S_0}(S_0'-\Phi_0'S_0'-\Psi_0'S_0'+
S_0'')\nonumber\\&-&\Phi_0'(\alpha k S_1+S_1')-\Psi_0'S_1'-
S_0'(\alpha k\Psi_1 +\Psi_1'+\Phi_1')\nonumber\\&+& \alpha
k S_1'+S_1''-S_0(\alpha k\Psi_1+\Psi_1')+\alpha k S_1+S_1'], \\
R_{11}^{(1)}(x)&=&\frac{e^{2(\Psi_0-\Phi_0+\tau)}}{\alpha^2}
[2(\Psi_1-\Phi_1)(\Psi_0'^2+\Psi_0''+\alpha\Psi_0'
\nonumber\\&-&\Phi_0'\Psi_0' +2\frac{\Psi_0' S_0'}{S_0})+
 2\alpha k\Psi_0'\Psi_1+2 \Psi_0'\Psi_1' +\alpha^2 k^2
\Psi_1\nonumber\\&+&2 \alpha k \Psi_1'+\Psi_1''+ \alpha^2 k
\Psi_1+\alpha \Psi_1'-\alpha k \Phi_0'\Psi_1-\Phi_0'\Psi_1'
\nonumber\\&-&\alpha k\Phi_1\Psi_0'-\Phi_1'\Psi_0'+ \frac{2}{S_0^2}
(\alpha k S_0 S_1\Psi_0'+ S_0 S_1'\Psi_0'\nonumber\\&+& \alpha k S_0
S_0'\Psi_1+ S_0 S_0'\Psi_1'-S_0'S_1\Psi_0')]-
e^{2(\frac{\tau}{\alpha}-x)}[-\Phi_1'\Psi_0'\nonumber\\
&+&\Phi_1''-\Phi_1'+2\Phi_0'\Phi_1'-\Phi_0'\Psi_1'-
\frac{2}{S_0}(S_1'\Psi_0'+S_0'\Psi_1'\nonumber\\&+&S_0\Psi_1'-
S_1''-S_1')+2\frac{S_1}{S_0^2} (S_0'\Psi_0'-S_0''-S_0')],\\
R_{22}^{(1)}(x)&=&-e^{-2\Psi_0}[2(S_1-S_0\Psi_1)
(\Phi_0'S_0'+\Phi_0'S_0-\Psi_0'S_0'-\Psi_0'S_0\nonumber\\
&+& S_0''+3 S_0'+S_0+\frac{S_0^2}{S_0})+\Phi_0'S_0 S_1'- \Psi_0'S_0
S_1'+2 S_0'S_1'\nonumber\\&+&3 S_0S_1'-
\Phi_0'S_0'S_1+\Psi_0'S_0'S_1-3 S_0'S_1+\Phi_1'S_0 S_0'
\nonumber\\&-&\Psi_0'S_0 S_0'+ S_0S_1'' -S_0''S_1+
\Phi_1'S_0^2-\Psi_1'S_0^2- 2\frac{S_0'^2
S_1}{S_0}]\nonumber\\&+&e^{2(\frac{\alpha-1}{\alpha}\tau+x-
\Phi_0)}\{ \alpha k(\alpha S_0 S_1+ \alpha k S_0 S_1+2S_0'S_1-
\Phi_0'S_0 S_1\nonumber\\&+&S_0(\Psi_0' S_1-\Phi_1 S_0'+\Psi_1
S_0'+2S_1'))+ \alpha S_0 S_1'+2 S_0' S_1'\nonumber\\
&+& S_0S_1'\Psi_0'-S_0 S_1'\Phi_0'+ S_0
S_0'\Psi_1'-S_0S_0'\Phi_1'+S_0 S_0''-\frac{S_0'^2
S_1}{S_0}\nonumber\\
&+&(S_1-2\Phi_1 S_0)(\alpha
S_0'+S_0'\Psi_0'-S_0'\Phi_0'+S_0'' +\frac{S_0'^2}{S_0})\},\\
R_{33}^{(1)}(x)&=&R_{22}^{(1)}(x)\sin^2\theta.
\end{eqnarray}
Now we can calculate the quantities $A_{ab}\equiv \phi_a\phi_b$
using Eq.(15) in Eqs.(10)-(14). Thus
\begin{equation}
A_{ab}(\tau,x,\epsilon)=\frac{1}{(-t)^2}\{{A_{ab}^{(0)}(x) +\epsilon
A_{ab}^{(1)}(x)e^{k\tau}+O(\epsilon^2)}\}.
\end{equation}
The perturbed part is given by
\begin{eqnarray}
A_{00}^{(1)}(x)&=&-\frac{e^{2\tau}}{\alpha}[4q(\alpha
k\phi_1+\phi_1')],\nonumber\\
A_{01}^{(1)}(x)&=&-e^{\frac{\alpha+1}{\alpha}\tau-x}
(2q\phi_1'),\nonumber\\
A_{11}^{(1)}(x)&=&0,\nonumber\\
A_{22}^{(1)}(x)&=&0, \nonumber\\
A_{33}^{(1)}(x)&=&0.
\end{eqnarray}
The linear perturbation equations can be written as
\begin{equation}
R_{ab}^{(1)}(x)=A_{ab}^{(1)}(x).
\end{equation}
Using Eqs.(27) and (28), it turns out that
\begin{eqnarray}
4\alpha q(\alpha k\phi_1+\phi_1')&=&\frac{3}{2}\alpha^2
k\Phi_1+\frac{3}{2}\alpha\Phi_1'+\alpha^2 k^2\Psi_1+ 2\alpha
k\Psi_1'\nonumber\\&+& \Psi_1''+\frac{2-\alpha}{4}
e^{\frac{1}{2}\alpha x}[(2k^2+2k+\frac{1}{2})
\alpha^2S_1\nonumber\\&+&2\alpha(2k+1)S_1'+2S_1''], \\
0&=&\Phi_1',\\
2\alpha q\phi_1'&=&(\alpha-2)(\alpha
k\Psi_1+\Psi_1')+\alpha\Phi_1'\nonumber\\&+&\frac{2-\alpha}{2}
e^{\frac{1}{2}\alpha x}[\alpha(2k+1)S_1\nonumber\\&+&(2\alpha
k+\alpha+2)S_1'+2S_1''],\\
0&=&2\alpha^2k(k-1)\Psi_1+2\alpha(2k-1)\Psi_1'\nonumber\\&+&
2\Psi_1''+\alpha(\alpha k\Phi_1+\Phi_1')+\alpha^2(\Psi_1-\Phi_1)\nonumber\\
&-&\frac{\alpha(2-\alpha)}{2}e^{\frac{1}{2}\alpha x}
[\alpha(2k+1)S_1+2S_1'],\\
0&=&2(2-\alpha)\Psi_1'+(\alpha-2)\Phi_1'+\Phi_1''\nonumber\\
&-&(2-\alpha)e^{\frac{1}{2}\alpha x}[\alpha
S_1+(\alpha+2)S_1'+2S_1''], \\
0&=&\alpha (\alpha k\Phi_1+\Phi_1')-\alpha(\alpha
k\Psi_1+\Psi_1')\nonumber\\&-&\alpha^2\Phi_1-\frac{2-\alpha}{2}
e^{\frac{1}{2}\alpha x}[\alpha^2k(2k-1)S_1\nonumber\\
&+&\alpha(4k-1)S_1'+S_1''], \\
0&=&(\alpha-2)(\Psi_1'-\Phi_1')-\Psi_1(\alpha^2-4\alpha\nonumber\\
&+&4)-\frac{2-\alpha}{2}e^{\frac{1}{2}\alpha x}
[(4-\alpha)S_1\nonumber\\&+&(6-\alpha)S_1'+2S_1''].
\end{eqnarray}

\section{Solutions of the Linear Perturbation Equations}

Now we solve system of the perturbed Eqs.(29)-(35). From Eq.(30), we
have
\begin{equation}
\Phi_1=b,
\end{equation}
where $b$ is an integration constant. Multiplying Eq.(35) by $2$,
adding in Eq.(33) and using Eq.(30), we obtain
\begin{equation}
(\alpha-2)\Psi_1+2e^{\frac{1}{2}\alpha x}[S_1+2S_1'+S_1'']=0.
\end{equation}
Using Eq.(36) in Eq.(34), it becomes
\begin{eqnarray}
0&=&\alpha^2(k-1)b-\alpha\Psi_1'-\alpha^2k\Psi_1\nonumber\\&+&
\frac{2-\alpha}{2}e^{\frac{1}{2}\alpha x}[\alpha^2k(2k-1)S_1+
\alpha(4k-1)S_1'+S_1''].
\end{eqnarray}
From Eq.(37), we have
\begin{equation}
\Psi_1=\frac{2}{2-\alpha}e^{\frac{1}{2}\alpha x} [S_1+2S_1'+S_1''].
\end{equation}
Eliminating $\Psi_1$ and $\Psi_1'$ from Eqs.(38) and (39), it turns
out that
\begin{equation}
A S_1+B S_1'+C S_1''+D S_1'''+Ee^{-\frac{1}{2}\alpha x}=0,
\end{equation}
where
\begin{eqnarray}
A&=&2\alpha^2(2k+1)-\alpha^2k(2-\alpha)^3(2k-1), \\
B&=&2\alpha(4\alpha k+2\alpha +2)-\alpha(4k-1)(2-\alpha)^3, \\
C&=&2\alpha (2\alpha k+\alpha +4)-(2-\alpha)^3, \\
D&=&4\alpha, \\
E&=&-2b\alpha^2(k-1)(2-\alpha).
\end{eqnarray}
It can be found that the solution with self-similarity of the second
kind is identical to the solution with first kind for the same type
of fluid. This means that the spacetime can have two different kinds
of self-similarities, i.e., there exist two vector fields
$\xi^\mu_{(1)},~\xi^\mu_{(2)}$, where $\xi^\mu_{(1)}$ describes
self-similarity of the first kind and $\xi^\mu_{(2)}$ of the second
kind. This happens when the spacetimes has high symmetry. As the
background solution with self-similarity of the second kind becomes
identical to the solution with self-similarity of the first kind, we
take $\alpha=1$. Thus Eqs.(41)-(45) become
\begin{eqnarray}
A&=&-2k^2+5k+2, \\
B&=&4k+9, \\
C&=&4k+9, \\
D&=&4, \\
E&=&-2bk+2b
\end{eqnarray}
and Eq.(40) takes the form
\begin{equation}
(-2k^2+5k+2)S_1+(4k+9)S_1'+(4k+9)
S_1''+4S_1'''+(-2bk+2b)e^{-\frac{1}{2}x}=0.
\end{equation}
This has the following solution
\begin{equation}
S_1(x)=c_1e^{Ux}-\frac{8b(k-1)}{8k^2-16k+3}e^{-\frac{1}{2}x},
\end{equation}
where $c_1$ is an integration constant and
\begin{eqnarray}
U&=&\frac{1}{12}P-\frac{B}{12A}+\frac{1}{12}A, \\
P&=&-9-4k, \\
B&=&27-24k-16k^2, \\
A&=&[297-756k+288k^2-64k^3+6\sqrt{3}\sqrt{\Delta}]^
{\frac{1}{3}}, \\
\Delta&=&999-4644k+6984k^2-3936k^3+1600k^4-512k^5.
\end{eqnarray}

\section{Boundary Conditions for the Perturbed Solutions}

In this section, we shall discuss some geometrical and physical
conditions [20] needed to be imposed for the spherically symmetry.
For gravitational collapse, we impose the following conditions:
\par\noindent
(i) There must exist a symmetry axis which can be expressed by
\begin{equation}
X\equiv\sqrt{|\xi^a_{(\theta)}\xi^b_{(\theta)}g_{ab}|} \rightarrow 0
\end{equation}
as $r \rightarrow 0$, where we have chosen the radial coordinate
such that the axis is located at $r=0$ and $\xi^a_{(\theta)}$ is a
Killing vector with a close orbit which is given by
$\xi^\alpha_{(\theta)}\partial_\alpha= \partial_\theta$.
\par\noindent
(ii) The spacetime near the symmetry axis is locally flat which can
be written by [19]
\begin{equation}
X_a X_b g^{ab}\rightarrow -1
\end{equation}
as $r \rightarrow 0$. It is mentioned here that solutions failing to
satisfy this condition may also be acceptable. Since we are mainly
interested in gravitational collapse, we assume that this condition
strictly holds at the beginning of the collapse so that we can be
sure that the singularity to be founded later on the axis is due to
the collapse.
\par\noindent
(iii) Closed timelike curves can be easily introduced in spacetimes
with spherical symmetry. For their absence the condition,
 \begin{equation}
\xi^a_{(\theta)}\xi^b_{(\theta)}g_{ab}<0,
\end{equation}
must hold in the whole spacetime. In addition to these conditions,
we also require that the spacetime is asymptotically flat in the
radial direction. For self-similar solutions, this condition cannot
be satisfied unless we restrict their validity only up to a maximal
radius say, $r=r_0(t)$, and join them with others in the region
$r>r_0(t)$ which are asymptotically flat as $r\rightarrow \infty$.
Here we simply assume that the self-similar solutions are valid in
the whole spacetime.

The boundary conditions at the event horizon $r_{AH}$, given by
Eq.(17), take the following form
\begin{equation}
r_{AH}=-t
\end{equation}
and the corresponding metric becomes
\begin{equation}
ds_{AH}^2=-4(-t)^2(d\theta^2+\sin^2\theta d\phi^2).
\end{equation}
This shows that the apparent horizon is singular only at $t=0$ and
the final state of the collapse is marginally naked singularity. Now
we discuss the following three cases:
\begin{eqnarray*}
(i)\quad\Delta>0,\quad (ii)\quad \Delta=0,\quad (iii)\quad\Delta<0.
\end{eqnarray*}
\textbf{Case (i):} In this case, using Eq.(52) in (39), we have
\begin{equation}
\Psi_1=-\frac{4b(k-1)}{8k^2-16k+3}+2(1+2U+U^2)C_1
e^{(U+\frac{1}{2})x}.
\end{equation}
For the first boundary condition, given by Eq.(58), we calculate the
quantity $\sqrt{X}=rS_1$ by using Eq.(8) in (52) so that
\begin{eqnarray}
S_1&=&c_1(\frac{r}{-t})^U-\frac{8b(k-1)}{8k^2-16k+3}
(\frac{r}{-t})^{-\frac{1}{2}},\\
rS_1&=&c_1\frac{r^{U+1}}{(-t)^U}-\frac{8b(k-1)}{8k^2-16k+3}
(-rt)^{\frac{1}{2}}.
\end{eqnarray}
For $rS_1\rightarrow 0$ as $r\rightarrow0$, all the exponents of $r$
must be greater than zero. It can be checked from Eqs.(53)-(56) that
$U+1$ is positive for $-\infty<k<1$ but turns out to be complex
otherwise. Thus the first condition is satisfied for $-\infty<k<1$
but is not fulfilled otherwise. Consequently, these perturbations
are limited by the boundary conditions for $1<k<\infty$.
\par\noindent
\textbf{Case (ii):} For $\Delta=0$, $k\rightarrow 1.70872$ along
with Eqs.(52) and (39), we have
\begin{equation}
\Psi_1=2.8876 b+0.2043 e^{-0.8196 x}.
\end{equation}
The boundary conditions, given by Eqs.(58)-(60), can be applied by
making use of Eq.(8) in Eq.(52)
\begin{eqnarray}
S_1&=&(\frac{r}{-t})^{-1.3196}+5.7752 b
(\frac{r}{-t})^{-\frac{1}{2}}, \\
r S_1&=&r^{-0.3196}(-t)^{1.3196}+5.7752 b
r^{\frac{1}{2}}(-t)^{\frac{1}{2}}.
\end{eqnarray}
Since all the exponents of $r$ are not positive, the condition given
by Eq.(58) is not satisfied. Again perturbations are limited by the
boundary conditions.
\par\noindent
\textbf{Case (iii):} Here we use Eq.(52) in Eq.(39) so that
\begin{equation}
\Psi_1=\frac{-4b(k-1)}{8k^2-16k+3}+(1+U_1)c_1 (\cos U_2+\iota\sin
U_2)e^{U_3x},
\end{equation}
where
\begin{eqnarray}
U_1&=&\frac{1}{12}(-9-4k)-\frac{27-24k-16k^2}
{12(297-756k+288k^2-64k^3-108\Delta)^{\frac{1}{6}}}
\nonumber\\&\times&\{\cos[\frac{1}{3}\arctan(\frac{6\sqrt{3}
\sqrt{-\Delta}}{297-756k+288k^2-64k^3})]\nonumber\\&-&
\iota\sin[\frac{1}{3}\arctan(\frac{6\sqrt{3}\sqrt{-\Delta}}
{297-756k+288k^2-64k^3})]\}\nonumber\\&+&\frac{1}{12}
(297-756k+288k^2-64k^3-108\Delta)^{\frac{1}{6}}\nonumber\\
&\times&\{\cos[\frac{1}{3}\arctan(\frac{6\sqrt{3}\sqrt{-\Delta}}
{297-756k+288k^2-64k^3})]\nonumber\\&-&\iota\sin
[\frac{1}{3}\arctan(\frac{6\sqrt{3}\sqrt{-\Delta}}
{297-756k+288k^2-64k^3})]\},
\end{eqnarray}
\begin{eqnarray}
U_2&=&\{[\frac{27-24k-16k^2}{12(297-756k+288k^2-64k^3-108\Delta)
^{\frac{1}{6}}}\nonumber\\&+&\frac{1}{12}(297-756k+288k^2-64k^3
-108\Delta)^{\frac{1}{6}}]\nonumber\\&\times&\sin[\frac{1}{3}\arctan
(\frac{6\sqrt{3}\sqrt{-\Delta}}{297-756k+288k^2-64k^3})]\}x, \\
U_3&=&\{\frac{1}{12}(-3-4k)-[\frac{27-24k-16k^2}
{12(297-756k+288k^2-64k^3-108\Delta)^{\frac{1}{6}}}\nonumber\\
&-&\frac{1}{12}(297-756k+288k^2-64k^3 -108\Delta)^{\frac{1}{6}}]
\nonumber\\&\times&\cos[\frac{1}{3}\arctan(\frac{6\sqrt{3}
\sqrt{-\Delta}}{297-756k+288k^2-64k^3})]\}x.
\end{eqnarray}
For the boundary condition, given by Eq.(58), we require that
$\sqrt{X}=rS_1$ and $S_1$ is real only if $c_1=0$. Thus we obtain
\begin{eqnarray}
S_1&=&-\frac{8b(k-1)}{8k^2-16k+3}
(\frac{r}{-t})^{-\frac{1}{2}},\\
rS_1&=&-\frac{8b(k-1)}{8k^2-16k+3} (-rt)^{\frac{1}{2}}.
\end{eqnarray}
Clearly, $rS_1\rightarrow 0$ as $r\rightarrow 0$, the condition,
given by Eq.(58), is satisfied only if
$k\neq1\pm\sqrt{\frac{5}{8}}$. For the second boundary condition,
given by Eq.(59), we have
\begin{eqnarray}
X_a X_b g^{ab}=-\frac{8192b^4(k-1)^4 r}{(8k^2-16k+3)^4} (b
r+\frac{\Psi_1}{t}).
\end{eqnarray}
Thus $X_a X_b g^{ab}\rightarrow 0$ as $r\rightarrow 0$.

\section{Concluding Remarks}

We have investigated the linear perturbations of the spherically
symmetric spacetimes with kinematic self-similarity of the second
kind. The self-similar variable and the background solution found
for the spherically symmetric spacetimes become identical to that of
the circularly symmetric metric [18]. However, the linearly
perturbed solution obtained for the spherically symmetric metric is
different from the exact solution for the circularly symmetric
metric. The boundary conditions for all possible values of $\Delta$
are discussed for this linearly perturbed solution.

For $\Delta>0$ and $\Delta=0$, the first boundary condition, given
by Eq.(58), is not satisfied. This shows that the perturbations are
limited by the boundary conditions in both these cases. For
$\Delta<0$, the boundary conditions are satisfied only for
$k\neq1\pm\sqrt{\frac{5}{8}}$ which admits both stable and unstable
modes for the perturbation. The stable modes are for $k<0$ and the
unstable modes are for $k>0$ but $k\neq1\pm\sqrt{\frac{5}{8}}$. The
unstable modes for the perturbation imply that it is not a critical
solution.

\vspace{1cm}

{\bf \large Acknowledgment}

\vspace{1cm}

The author would like to thank referee for the constructive
comments.

\vspace{1cm}

{\bf \large References}

\begin{description}

\item{[1]} Choptuik, M.W.: Phys. Rev. Lett.
\textbf{70}(1993)9.

\item{[2]} Choptuik, M.W., Hirchmann, E.W., Liebling, S.L.
 and Pretorius, S.: Phys. Rev. \textbf{D68}(2003)044007.

\item{[3]} Hara, T., Koike, T. and Adachi, S.: Phys. Rev.
\textbf{D59}(1999)104008.

\item{[4]} Garfinkle, D.: Phys. Rev. \textbf{D63}(2001)0044007.

\item{[5]} Pretorius, F. and Choptuik, M.W.: Phys. Rev.
\textbf{D62}(2000)124012.

\item{[6]} Garfinkle, D. and Gundlach, C.: Phys. Rev.
\textbf{D66}(2002)044015.

\item{[7]} Chandrasekhar, S.: \textit{The Mathematical Theory of Black Holes}
(Clarendon Press, Oxford University Press, Oxford, 1983).

\item{[8]} Hirchmann, E.W., Wang, A. and Wu, Y.: Class.
Quantum Grav. \textbf{21}(2004)1791.

\item{[9]} Clement, G. and Fabbri, A.: Class. Quantum Grav.
\textbf{18}(2001)3665; Nucl. Phys. \textbf{B630}(2002)269.

\item{[10]} Cavaglia, M., Clement, G. and Fabbri, A.: Phys. Rev.
\textbf{D70}(2004)044010.

\item{[11]} Miguelote, A.Y., Tomimura, N.A. and Wang, A.: Gen.
Rel. Grav. \textbf{36}(2004)1883; arXiv: gr-qc/0505062.

\item{[12]} Frolov, A.V.: Phys. Rev. \textbf{D56}(1997)6433; ibid
\textbf{59}(1999)104011.

\item{[13]} Brandt, C.F.C., Lin, L.M., Villas da Rocha, J.F.
 and Wang, A.: Int. J. Mod. Phys. \textbf{D11}(2002)155.

\item{[14]} Benoit, P.M. and Coley, A.A.: Class. Quantum Grav.
\textbf{15}(1998)2397.

\item{[15]} Wang, A.: Phys. Rev. \textbf{D68}(2003)064006.

\item{[16]} Wang, A., Wu, Y. and Wu, Z.C.: Gen. Rel. Grav.
\textbf{36}(2004)1225.

\item{[17]} Chan, R., da Silva, M.F.A., Villas da Rocha,
J.F. and Wang, A.: Int. J. Mod. Phys. \textbf{D14}(2005)1049.

\item{[18]} Chan, R., da Silva, M.F.A. and Villas da Rocha,
J.F.: Int. J. Mod. Phys. \textbf{D14}(2005)1725.

\item{[19]} Stephani, H., Kramer, D., MacCallum, M., Hoenselaers, C. and Herlt, E.:
\textit{Exact Solutions of Einstein Field Equations} (Cambridge
University Press, 2003).

\item{[20]} Carlip, S.: \textit{Quantum Gravity in (2+1)-Dimensions}
(Cambridge University Press, Cambridge, 1998).

\end{description}

\end{document}